\documentclass[sigconf]{acmart}

\usepackage{graphicx}
\usepackage{subcaption}
\usepackage{algorithmic}
\usepackage{algorithm}
\usepackage{dsfont}
\usepackage{balance} 

\AtBeginDocument{%
  \providecommand\BibTeX{{%
    \normalfont B\kern-0.5em{\scshape i\kern-0.25em b}\kern-0.8em\TeX}}}

\copyrightyear{2022} 
\acmYear{2022} 
\setcopyright{acmlicensed}
\acmConference[WWW '22]
{Proceedings of the ACM Web Conference 2022}{April 25--29, 2022}{Virtual Event, Lyon, France}
\acmBooktitle{Proceedings of the ACM Web Conference 2022 (WWW '22), April 25--29, 2022, Virtual Event, Lyon, France}
\acmPrice{15.00}
\acmDOI{10.1145/3485447.3512194}
\acmISBN{978-1-4503-9096-5/22/04}

\begin{document}

\title{Generating Simple Directed Social Network Graphs for Information Spreading}

\author{Christoph Schweimer}
\email{cschweimer@know-center.at}
\affiliation{%
  \institution{Know-Center GmbH}
  \city{Graz}
  \country{Austria}
}

\author{Christine Gfrerer}
\email{cgfrerer@cs.sbg.ac.at}
\affiliation{%
  \institution{University of Salzburg}
  \city{Salzburg}
  \country{Austria}
}

\author{Florian Lugstein}
\email{flugstein@cs.sbg.ac.at}
\affiliation{%
  \institution{University of Salzburg}
  \city{Salzburg}
  \country{Austria}
}

\author{David Pape}
\email{david.pape@stud.sbg.ac.at}
\affiliation{%
  \institution{University of Salzburg}
  \city{Salzburg}
  \country{Austria}
}

\author{Jan A. Velimsky}
\email{jan.velimsky@gsi.uni-muenchen.de}
\affiliation{%
  \institution{DIALOGIK gGmbH}
  \city{Stuttgart}
  \country{Germany}
}

\author{Robert Els\"asser}
\email{elsa@cs.sbg.ac.at}
\affiliation{%
  \institution{University of Salzburg}
  \city{Salzburg}
  \country{Austria}
}

\author{Bernhard C. Geiger}
\email{geiger@ieee.org}
\affiliation{%
  \institution{Know-Center GmbH}
  \city{Graz}
  \country{Austria}
}

\renewcommand{\shortauthors}{Schweimer, et al.}

\begin{abstract}
Online social networks are a dominant medium in everyday life to stay in contact with friends and to share information. In Twitter, users can connect with other users by following them, who in turn can follow back. In recent years, researchers studied several properties of social networks and designed random graph models to describe them. Many of these approaches either focus on the generation of undirected graphs or on the creation of directed graphs without modeling the dependencies between reciprocal (i.e., two directed edges of opposite direction between two nodes) and directed edges. We propose an approach to generate directed social network graphs that creates reciprocal and directed edges and considers the correlation between the respective degree sequences.

Our model relies on crawled directed graphs in Twitter, on which information w.r.t.\ a topic is exchanged or disseminated. While these graphs exhibit a high clustering coefficient and small average distances between random node pairs (which is typical in real-world networks), their degree sequences seem to follow a $\chi^2$-distribution rather than power law. To achieve high clustering coefficients, we apply an edge rewiring procedure that preserves the node degrees.

We compare the crawled and the created graphs, and simulate certain algorithms for information dissemination and epidemic spreading on them. The results show that the created graphs exhibit very similar topological and algorithmic properties as the real-world graphs, providing evidence that they can be used as surrogates in social network analysis. Furthermore, our model is highly scalable, which enables us to create graphs of arbitrary size with almost the same properties as the corresponding real-world networks.
\end{abstract}



\maketitle

\section{Introduction}

Online social networks, such as Facebook and Twitter have become leading platforms for people to stay in touch with others, communicate and exchange news and information. Since their emergence, researchers have studied the structure of the network graphs, utilised various topological features to describe and analyze them, and simulated the spread of messages therein. Some platforms offer the possibility to obtain the social network graph or parts of it, but the process of crawling these graphs is often very time-consuming, especially if they consist of millions of nodes and edges. Additionally, the information in these graphs needs to be handled carefully due to the sensitive nature of the user data and data protection regulations. Thus, there is a need to model and synthetically generate realistic social network graphs. The Web Conference, which is the premier international event focusing on the study of the World Wide Web, devotes one of its major tracks to the ``investigation of graph-based techniques for social networks'' and the development of ``new theories, models, and algorithms to make these systems more effective and efficient''\footnote{See \url{https://www2022.thewebconf.org/cfp/research/sna/}}.

Random graph generation dates back way before the invention of online social networks. Most approaches either create undirected graphs or directed graphs without specifically modeling reciprocal edges (two directed edges of opposite direction between two nodes).
In a recent seminal paper \cite{CDGQ+21}, the authors analyzed the spread of information in online social networks and concluded that ``the aggregation of users in homophilic clusters dominate online interactions on Facebook and Twitter''. Motivated by the connection between interaction networks and the spread of information in social networks, we analyze real-world graphs on which information w.r.t.\ certain topics are exchanged and disseminated.
We develop a synthetic graph generation method for such interaction networks that not only creates a graph with reciprocal and directed edges, but also takes the correlation between reciprocal and directed node degrees into account via Copulas. Nodes are then connected by edges according to their sampled degrees and edges are rewired to increase the average clustering coefficient to a realistic level.

We applied our method to $14$ crawled subgraphs of the Twitter follower graph of different sizes and structures, and created graphs with the same number of nodes. Results show that our approach creates graphs that match several topological features of the crawled network graphs, and that behave similarly when applying certain processes, e.g., information dissemination. Our method can also create graphs of arbitrary size with various degree distributions.

The paper is structured as follows. In Section \ref{RELATED_WORK} we give a short summary about different graph generation methods and in Section \ref{CRAWLED_NETWORKS} we present the crawled subgraphs. In Section \ref{METHOD} we describe our graph creation method, and we present our results in Section \ref{RESULTS}. The discussion in Section \ref{DISCUSSION} concludes the paper.

\section{Related Work}

\label{RELATED_WORK}

One of the first random graph generation models is the \textit{Erd\H{o}s-R\'enyi} model \cite{Erdos1959}, where for given numbers of nodes and edges, one of the possible graphs is chosen uniformly at random. An alternative version is presented in \cite{Gilbert1959}, where nodes are connected independently according to a predefined probability. Erd\H{o}s-R\'enyi graphs have small clustering coefficients, but form the basis for random graph generation. 
A model with high clustering coefficient was designed by Watts and Strogatz \cite{Watts1998}. In this model, the graph starts as a regular ring lattice (i.e., every node is connected to its closest $k$ neighbors on a ring, for some constant $k$) and edges are iteratively rewired.
A different class of graph generation models is based on growth and the Preferential Attachment (PA) mechanism. The initial graph in the \textit{Barab\'asi-Albert} model \cite{Barabasi1999} contains $m_0$ nodes. In each step a new node is added to the graph (growth), which connects to $m \leq m_0$ nodes chosen with probability proportional to their current degree (PA). This method has later been extended to directed graphs, see e.g., \cite{Bollobas2003}.

Other random graph generation methods are based on establishing connections between nodes according to node degree sequences. One of these methods is the \textit{Configuration} model \cite{Bender1978, Bollobas1980} whose functionality is described in \cite{Britton2006} and \cite{Chen2013}.
The concept for undirected graphs is to sample node degrees from a distribution, create stubs for each node and randomly connect two of them until no stubs are left. The sampled node degrees are met exactly in the generated graphs, but they often contain self-loops (a node connected to itself) and parallel edges (two or more edges between two nodes). To create a simple graph (no self-loops and only one edge between nodes) with this method, alternatives are presented in \cite{Britton2006} and \cite{Chen2013}. 
The authors of \cite{Chen2013} additionally introduce a directed version of the Configuration model, where the sum of the in-degrees and the sum of the out-degrees must be equal such that all stubs can connect. 
Similar to the idea of the Configuration model, but with a probabilistic approach, is the \textit{Chung-Lu} model \cite{Chung2002}. Each node pair is connected with a probability proportional to the product of their respective degrees. 
The sampled degree sequences are not met exactly, but the created graphs do not contain parallel edges. A directed version of the Chung-Lu model is presented in \cite{Durak2013}. We should also mention the block two-level Erd\H{o}s-R\'enyi model \cite{SKP12,KPPS14}, which includes a community structure into the graphs.
First, it creates communities using Erd\H{o}s-R\'enyi graphs (called affinity blocks), and then edges are generated between communities using a variation of the Chung-Lu model. 
A structured summary of additional graph generation approaches can be found in \cite{Bonifati2020}, where the authors provide an extensive survey on state of the art graph generators, discuss their strengths, weaknesses and open challenges.

Various approaches for directed graphs exist, but most of them do not explicitly model reciprocal edges in directed graphs, even though real-world graphs often exhibit a high amount of reciprocity \cite{Durak2013, Kwak2010}. 
The authors of \cite{Durak2013}, for example, extend the Chung-Lu model and provide a powerful baseline model to match the reciprocal, in- and out-degree distributions of a graph; however, their model does not take the correlation between the node degrees into account.
Also, it is known that many social network graphs exhibit a high average clustering coefficient (CC) \cite{Myers2014,Ahn2007,Mislove2007}, which has to be taken into account when generating random graphs with corresponding topological properties.
\cite{Newman2009} propose a modified Configuration model to create undirected graphs with high clustering by not only specifying the number of edges per node, but also the number of triangles per node. Extending this approach to directed graphs does not seem straightforward since for every triangle, the type of edge between two nodes would also need to be determined.

Alternatively, the technique of \textit{edge rewiring} (also called edge switching) was developed to increase the CC in a graph.
The authors of \cite{Alstott2016} introduce various rewiring algorithms where one edge is switched per iteration (e.g., edge $(i-j)$ is replaced by $(j-k)$), which alters the node degrees.
In another approach, two non-adjacent edges are rewired at a time and the new graph is accepted if the CC has increased.
In \cite{Bansal2009} and \cite{Guo2009}, the idea is to randomly select a node $x$ in the graph, two unconnected neighbors of $x$, $y_1$ and $y_2$, and two unconnected second degree neighbors of $x$, $z_1$ and $z_2$, with edges $(y_1 - z_1), (y_2 - z_2)$. 
Changing these edges to $(y_1 - y_2)$ and $(z_1 - z_2)$ increases the CC of node $x$ and preserves the degree sequences. 
The authors of \cite{Tabourier2011} propose switching more than two edges at a time by applying a random permutation on the destination nodes, thus preserving the degree sequences. The new graph is accepted if it fulfills predefined graph constraints (e.g., graph is simple). 

Our graph creation method resembles the idea of \cite{Durak2013} by treating reciprocal and directed edges separately, but additionally considers the rank correlation between the node degrees when sampling them. Nodes are then connected according to the principle of the Chung-Lu model \cite{Chung2002}, but avoiding self-loops or parallel edges. To increase the clustering coefficient of the created graph, the rewiring procedure in \cite{Bansal2009} is extended to directed graphs. Instead of randomly selecting one node in each iteration and aiming for a predefined CC, we apply several rewiring iterations on every node, depending on its degree, up to a certain node degree.
Considering only a subset of options for rewiring induces a bias into our proposed graph generation method. \cite{Tao2016,McDonald2007} present and discuss methods for uniform random graph generation (also by utilising higher-order edge switches), which usually have a very high guaranteed running time \cite{Greenhill15,Fosdick18}. 
We therefore accept this induced bias and favor increasing the CC of smaller degree nodes, which also tend to have a higher CC in social networks, cf. \cite{Mislove2007,Myers2014}.

\section{Dataset Description}
\label{CRAWLED_NETWORKS}

Our work is based on Twitter follower networks, which is motivated by the results of \cite{CDGQ+21} regarding the interaction networks, on which information about specific topics is exchanged or disseminated. We acquired six different Tweet datasets from the Twitter Historical Power Track. Tweets were collected in a time frames of 41 days between August 2019 and March 2020, which we believe results in sufficiently complete graphs of users for the respective topics. The topics include vegan diet, social distance, two soccer clubs in Germany and two political parties in Austria. Eight datasets were acquired through Premium Search API of Twitter, between August 2021 and October 2021, and the time frames are 4, 10 or 14 days depending on the topic, allowing us to investigate networks over shorter time periods. These topics cover mainly digitization trends and computer science fields. For each of the 14 topics, we extracted the users who actively posted, retweeted or quoted, and then crawled their followers using the Twitter API. The lists of followers were truncated to only keep the follower relationships between the users of a topic. Finally, user IDs were anonymized. 

The follower network graphs $G = (V,E)$ are directed, where each node $v \in V$ corresponds to a user and each edge $e \in E$ represents a follower relationship between two users. In Twitter a user $v_1$ can follow another user $v_2$ which connects them by a directed edge ($v_1 \rightarrow v_2$) and if two users follow each other, they are connected by two directed edges of opposite direction, which we will treat as one reciprocal edge ($v_1-v_2$).
For each node in the $14$ graphs ($G_1$ through $G_{14}$), we determined the reciprocal degree (number of neighbors that are connected via a reciprocal edge), the in-degree (total incoming edges minus reciprocal degree), the out-degree (total outgoing edges minus reciprocal degree) and the total degree (number of neighbors), which is the sum of the three node degrees above.
We computed \textit{Spearman's rank correlation coefficient} $\rho$ between the degrees and also several topological graph features, studied and defined in \cite{Myers2014,Ahn2007,Mislove2007} to characterize the crawled graphs: The sizes of the largest strongly connected component (LSCC) and the largest weakly connected component (LWCC), the density, the average shortest path length (ASPL), the diameter and the average clustering coefficient (CC), all related to the LWCC. The statistics of $5$ of the crawled graphs are listed in the first line of the respective cells in Table \ref{Results_5}, and in Table \ref{Results_all} in Annex \ref{Additional_results} for the remaining $9$ graphs. Features related to the LWCC have an asterisk (e.g., Density* for the density in the LWCC). As typical for social networks \cite{Myers2014,Ahn2007,Mislove2007}, the crawled graphs are characterised by small shortest path lengths, huge connected components, small diameter and high average CC.

\section{Graph Creation Method}

\label{METHOD}

In this section we introduce the method to synthetically create realistic social network graphs with reciprocal and directed edges\footnote{For the code see \url{https://gitlab.com/eu_hidalgo/use_cases/-/tree/master/social_network/network_generation}}. Our goal is to create graphs with the same number of nodes as the crawled ones, a comparable density and a similar average CC in the LWCC. We present the information extracted from the crawled graphs to generate node degrees and establish edges between the nodes without creating self-loops or parallel edges. An edge rewiring procedure concludes the approach.
 
\subsection{Node Degree Sampling}

\label{NODE_DEGREE_GENERATION}

\begin{figure}[ht]
\includegraphics[width=0.9\linewidth]{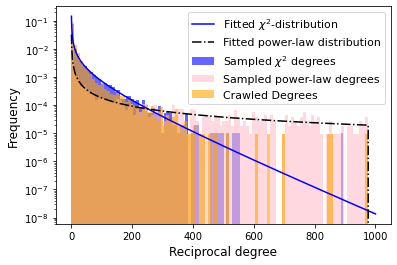}
\caption{Comparison of the fitted power-law and the fitted $\chi^2$-distribution for the reciprocal degree in $G_1$}
\label{Hist_1}
\end{figure}

For the synthetic graph creation, we sample the reciprocal degree, the in-degree and the out-degree for each of the $n=|V|$ nodes from fitted distributions. Most literature suggests that the degrees of a social network graph follow a power-law or a log-normal distribution \cite{Myers2014,Ahn2007,Mislove2007}. We performed Kolmogorov-Smirnov tests to fit probability distributions to the three node degrees, but the returned $p$-values were close to $0$ for each distribution. By testing several distributions, we found that the degree sequences of the crawled and subsequently truncated Twitter subgraphs considered in this paper are best modeled by $\chi^2$-distributions, if hyperparameters were selected such that mean and standard deviation agrees with those of the crawled graphs.
The log-scale histogram in Figure \ref{Hist_1} shows that there are only a few nodes with a high reciprocal degree, e.g., $>500$, in the crawled graph $G_1$ (orange bins). A similar behavior regarding the node degree distribution was observed for the in- and out-degree in $G_1$, and also for the degrees in the other graphs. Even though the fitted $\chi^2$-distribution (blue bins) tends to generate a smaller amount of high-degree nodes, it fits the degree distribution of the crawled graph well w.r.t.\ the quartiles, the mean degree and the standard deviation, whereas the power-law distribution (pink bins) overproduces nodes with a high degree.

Sampling the reciprocal, in- and out-degree for the nodes of directed graphs independently results in nodes having unrealistic joint degrees such as high in- and out-degrees but a reciprocal degree of 0, which we did not observe in the crawled graphs. Indeed, there is a high rank correlation (Spearman's rank correlation coefficient $\rho$) between the reciprocal, in- and out-degree in the crawled graphs (see the first lines of the respective cells in Table \ref{Results_5} and Table \ref{Results_all} in Annex \ref{Additional_results}). 
To prescribe the rank correlation between sampled node degrees, we make use of the \textit{NORTA} principle (normal to anything) \cite{Cario1997}, where multivariate normally distributed data is sampled and transformed into the desired joint distribution in two steps.
First, we fit $\chi^2$-distributions to the reciprocal, in- and out-degree and compute the rank correlation coefficients $\rho_1 = \rho \text{(reciprocal, in)}$, $\rho_2 = \rho \text{(reciprocal, out)}$, $\rho_3 = \rho \text{(in, out)}$ between the three degrees.
Next, we map $\rho_1, \rho_2, \rho_3$ into the linear correlation coefficients $R_1, R_2, R_3$ with $\displaystyle R_k = 2 \cdot \sin(\rho_k \cdot \pi/6)$ \cite{Pearson1907}, $k=1,2,3$, to sample $n$ times from a multivariate Gaussian distribution with mean vector $\mu = (0,0,0)$ and covariance matrix
$$\Sigma = \begin{pmatrix}
 1 & R_1 & R_2 \\
 R_1 & 1 & R_3 \\
 R_2 & R_3 & 1 \\
\end{pmatrix}.$$ 
Applying the cumulative distribution function (CDF) transforms the sampled multivariate data into the $]0,1[^3$-space, where each dimension of the data approximately follows the uniform distribution $\mathcal{U}_{]0,1[}$. The joint distribution $\mathcal{U}_{]0,1[}^3$ is known as a \textit{Copula} \cite{Nelsen2006}. To receive the synthetic node degrees from the Copula, we use the inversion method \cite{Devroye1986}. We apply the inverse of the CDF of the fitted $\chi^2$-distributions to the respective dimensions of the Copula data and round the results to integers. 
This returns the synthetically generated reciprocal, in- and out-degree sequences $\mathbb{R}, \mathbb{I}, \mathbb{O}$ of length $n$, which have approximately the same rank correlation as the node degree sequences of the crawled graphs. We interpret the degree sequences as indexed multisets, i.e., the degrees of each node are at the same position in $\mathbb{R}, \mathbb{I}, \mathbb{O}$, such that we can use set notation.

\subsection{Connecting nodes via edges}

\label{CONNECTING}

We now sample edges by computing the connection probability of two nodes based on their generated degrees while avoiding self-loops and parallel edges. First we sample reciprocal edges and based on these edges, we sample directed edges. Our approach resembles the Chung-Lu model \cite{Chung2002,Durak2013}, but results in a simple graph. 

The method requires as input an empty graph $G = (V,E)$ that contains $n = |V|$ numbered nodes $V = \{1, \ldots, n\}$ and the generated reciprocal, in- and out-degree sequences $\mathbb{R}, \mathbb{I}, \mathbb{O}$. We determine the expected reciprocal degree sum $r = \sum_{i=1}^n \mathbb{R}(i)$ and the expected number of directed edges $d = \frac{1}{2} \cdot \sum_{i=1}^n (\mathbb{O}(i) + \mathbb{I}(i))$. 
Edge sampling is a Bernoulli experiment, only using the nodes with respective degrees unequal to zero. I.e., for a node pair $(i,j)$ we place a reciprocal edge $(i-j)$ (implemented as two directed edges $(i \rightarrow j)$ and $(j \rightarrow i)$) with probability proportional to $\mathbb{R}(i) \cdot \mathbb{R}(j)$, and a directed edge $(i\to j)$ with probability proportional to $\mathbb{O}(i) \cdot \mathbb{I}(j)$.

To avoid self-loops, before sampling reciprocal edges, we compute the probability sum of the reciprocal self-loops and uniformly add it to the connection probabilities of the possible reciprocal edges (where $\mathbb{R}(i) \neq 0$ and $\mathbb{R}(j) \neq 0$).
Before sampling directed edges, we first inspect if a directed edge $(i \rightarrow j)$ has already been sampled as a reciprocal edge. 
If this is the case, we add its probability to a reservoir of already sampled potential directed edges (the same is done for the edge $(j \rightarrow i)$) which will be uniformly added to the possible directed edges ($\mathbb{O}(i) \neq 0$ and $\mathbb{I}(j) \neq 0$). For directed self-loops, we apply the same technique as for reciprocal self-loops. 

\subsection{Edge Rewiring}

\label{REWIRING}

\begin{algorithm}[t]
\small
\caption{Edge rewiring}
\label{Algorithm_rewiring}
\begin{algorithmic}[1]

\REQUIRE Directed graph $G = (V,E)$ with $V= \{1, \ldots, n\}$

\STATE $t = 95-$percentile $\deg(V)$
\STATE $V' = V \setminus (\{\deg(V) < 2\} \cup \{\deg(V) > t\})$
\STATE $m = \text{median}(\deg(V'))$


\STATE \textbf{for all} $x \in V'$

\STATE \hspace*{3mm} \textbf{if} $\deg(x) \leq m$
\STATE \hspace*{6mm} $\alpha = \deg(x) \cdot (\deg(x)-1) / 2 $
\STATE \hspace*{3mm} \textbf{else}
\STATE \hspace*{6mm} $\alpha = \lceil \deg(x) \cdot (\deg(x)-1) / 2 \cdot 0.6 \rceil$
\STATE \hspace*{3mm} \textbf{for} $\alpha$-times \label{alpha}
\STATE \hspace*{6mm} Randomly select $y_1\in ne(x)$ and $y_2 \in ne(x) \setminus y_1$ \label{Choose}
\STATE \hspace*{6mm} \textbf{if} $(y_1 \rightarrow y_2) \notin E$ and $(y_2 \rightarrow y_1) \notin E$
\STATE \hspace*{9mm} \textbf{for} $10$-times \label{10-times}
\STATE \hspace*{12mm} Randomly select $z_1 \in ne(y_1) \setminus ne(y_2)$
\STATE \hspace*{12mm} Randomly select $z_2 \in ne(y_2) \setminus ne(y_1)$
\STATE \hspace*{12mm} \textbf{if} $(z_1 \rightarrow z_2) \notin E$ and $(z_2 \rightarrow z_1) \notin E$
\STATE \hspace*{15mm} Rewire $(y_1,y_2,z_1,z_2)$ (see Algorithm \ref{Algorithm_rewiring_step})
\STATE \hspace*{15mm} \textbf{break}

\end{algorithmic}
\end{algorithm}

The degree distributions of the nodes in the created graphs match the desired degree distributions well, but the average CC in the LWCC of the created graphs is lower than in the crawled graphs, see Table \ref{Results_5_intermediary} and Table \ref{Results_before_rewiring} in Appendix \ref{Additional_results} for comparison. To increase the CC, we utilise the technique of edge rewiring. We are following the approaches presented in \cite{Bansal2009} and \cite{Guo2009}, but instead of randomly selecting one node per iteration, we consider all nodes up to a certain total degree (95-percentile of all node degrees) for the rewiring procedure. For each node $x \in V$, we try to connect a fraction of its first degree neighbors, depending on the degree of $x$. We do not compute the CC after every rewiring iteration since this is time-consuming, and we do not aim for a target CC. 
The pseudocode for the edge rewiring procedure is depicted in Algorithm \ref{Algorithm_rewiring}.

Since nodes with a high degree exhibit a smaller average CC \cite{Myers2014,Mislove2007}, we do not apply the rewiring procedure on the nodes with the highest degrees. Experiments have shown that it suffices to only consider those 95 percent of all nodes with the lowest total degree. For ease of notation, we use $\deg(V)$ to denote the indexed multiset of node degrees in $V$. From this set of nodes, we remove all nodes in the upper 95-percentile of node degrees and all nodes with a degree smaller than 2 since the rewiring procedure cannot be applied for them. The resulting set of nodes is denoted as $V'$.

We look at each node $x \in V'$ in ascending order of its node ID, i.e., potentially starting with node 1, and iterate multiple times over each node, depending on its degree $\deg(x)$, to apply the rewiring procedure.
Since nodes with a high degree tend to have a smaller average CC, we sample $60\%$ of first degree neighbor pairs for nodes with a degree above the median degree $m$ in $V'$ and all first degree neighbor pairs otherwise. One rewiring iteration (line \ref{alpha} in Algorithm \ref{Algorithm_rewiring}) for a node $x \in V'$ starts with the random selection of two distinct first degree neighbors $y_1 \in ne(x)$ and $y_2 \in ne(x)$. 
If they are connected, they are not further considered in the rewiring process and a new first degree neighbor pair is selected. If they are not connected, we look at their neighbors, denoted as $ne(y_1)$ and $ne(y_2)$ and remove the nodes that are connected to both, $y_1$ and $y_2$. We additionally remove all nodes that have a smaller node ID than $x$ to not disrupt the already investigated nodes too much. 
Next, we randomly sample two nodes $z_1 \in ne(y_1)$ and $z_2 \in ne(y_2)$. Rewiring can only happen if i) $z_1$ and $z_2$ are not connected and ii) to preserve the node degrees, if $y_1, y_2$ are connected to $z_1, z_2$ via reciprocal edges or if they are connected via directed edges with opposite directions
(see Algorithm \ref{Algorithm_rewiring_step} in Appendix \ref{Additional_algorithm}). As these conditions may not always be fulfilled, we sample the second degree neighbors a maximum of 10 times (line \ref{10-times} in Algorithm \ref{Algorithm_rewiring}).

\section{Results}
\label{RESULTS}

\begin{table}
\small
\caption{Topological features of $5$ crawled graphs (line 1) and the corresponding created graphs (line 2)}
\centering 
 \begin{tabular}{l c c c c c} 
\toprule
  & $G_1$ & $G_2$ & $G_3$ & $G_4$ & $G_5$\\ 
 \midrule
 Nodes & 11,015 & 21,291 & 50,133 & 459 & 3,580 \\ 
 \midrule
 
 Edges & 377,457 & 2,570,452 & 4,832,226 & 5,435 & 54,735\\ 
  & 381,627 & 2,527,541 & 5,049,608 & 5,499 & 57,249\\ 
 \midrule
 
 Density & 0.0031 & 0.0057 & 0.0019 & 0.0259 & 0.0043\\
  & 0.0031 & 0.0056 & 0.0020 & 0.0262 & 0.0045\\ 
  \midrule
 
 LSCC & 9,347 & 19,237 & 43,461 & 361 & 2,314\\ 
  & 8,182 & 19,296 & 41,805 & 390 & 1,999\\  
 \midrule
 
 LWCC & 10,931 & 21,281 & 49,999 & 452 & 3,570\\ 
  & 10,166 & 21,148 & 48,607 & 446 & 3,354\\ 
 \midrule

 Density* & 0.0032 & 0.0057 & 0.0019 & 0.0266  & 0.0043\\ 
  & 0.0037 & 0.0057 & 0.0021 & 0.0277 & 0.0051\\ 
  \midrule
 
 ASPL* & 2.97 & 2.69 & 2.75 & 2.40 & 2.34\\ 
  & 2.59 & 2.70 & 2.79 & 2.48 & 1.99\\ 
 \midrule
 
 Diameter* & 11 & 9 & 10 & 8 & 11\\ 
  & 7 & 6 & 7 & 5 & 6\\ 
 \midrule
 
 Average CC* & 0.201 & 0.285 & 0.198 & 0.354 & 0.277\\
  & 0.228 & 0.291 & 0.222 & 0.300 & 0.187\\ 
 \midrule
 
  Runtime & 1h & 31h & 110h & 1min & 3min \\
\midrule
 
 $\rho_1$ & 0.540 & 0.630 & 0.616 & 0.465 & 0.407\\
  & 0.470 & 0.604 & 0.579 & 0.259 & 0.319\\
 \midrule
 
 $\rho_2$ & 0.612 & 0.697 & 0.753 & 0.606 & 0.547\\
  & 0.528 & 0.660 & 0.696 & 0.419 & 0.411\\
 \midrule
 
 $\rho_3$ & 0.284 & 0.395 & 0.397 & 0.240 & 0.250\\
  & 0.234 & 0.369 & 0.371 & 0.140 & 0.149\\
 \bottomrule
\end{tabular}
\label{Results_5}
\end{table}

\subsection{Topological Features}

To evaluate the graph creation method, we aimed to replicate the $14$ crawled graphs described in Section \ref{CRAWLED_NETWORKS}. For the topological features, the main goal was to create graphs with the same number of nodes, a similar density and a comparable average CC in the LWCC.
The topological features of $5$ crawled graphs and the corresponding synthetically created graphs are listed in Table \ref{Results_5} (the remaining ones are listed in Table \ref{Results_all} in Annex \ref{Additional_results}). 

The results show that our approach creates graphs with a similar size as the crawled graphs. The number of edges in the graphs are in a similar range and the density only starts to differ in the fourth decimal place. Due to the assignment of degrees to nodes according to rank correlations and the probabilistic sampling of edges, some nodes have a total degree of 0. Thus, the sizes of the largest connected components in the created graphs have the tendency to be smaller than in the crawled ones.

We observe that the average shortest path length (ASPL) and the diameter also tend to be slightly smaller in the created graphs. Since nodes connect to each other with a probability proportional to their degrees, low-degree nodes are more likely to connect to nodes with a high degree than to nodes with low degree, leading to short path lengths between nodes and a small diameter in the created graphs. In the crawled graphs however, we observed nodes with a degree of one, known as \textit{leaves}, that are connected to another node with a small degree. These leaves have a high shortest path length to each other causing a higher diameter.

In the crawled graphs, the average CC lies between 0.20 and 0.35, while in created graphs the span reaches from 0.15 to 0.38. Since we are not aiming for a fixed CC, these fluctuations are expected. Both for crawled and created graphs, a higher density in the LWCC (Density*) tends to correlate with a higher average CC. Even though the difference between the crawled and the created graph in $G_7$ is above 0.1 (see Table \ref{Results_all} in Annex \ref{Additional_results}), the result is still acceptable, since the created graph exhibits a reasonable average CC* ($0.150$).

The last three cells of Table \ref{Results_5} (and Table \ref{Results_all}) contain the rank correlations between reciprocal, in- and out-degree, where we see that the correlation in the created graphs (line 2 in the cells) is lower than in the crawled ones (line 1 in the cells). This difference is caused by rounding the sampled data to integers (Section \ref{NODE_DEGREE_GENERATION}) and by probabilistic edge sampling (Section \ref{CONNECTING}), where the node degrees are not replicated exactly. The second offset could be avoided by applying an adjusted version of the Configuration model.

We also analyzed the community structure of the crawled and the created graphs. When used for comparing random graph models to real-world networks, usually disjoint communities have been considered (see e.g., \cite{KPPS14}). Here we focus on overlapping communities, which often can be observed in social networks (i.e., each user is usually part of several communities). To obtain these overlapping communities, we cluster the neighbourhood of each node $v$, using the parallelized Louvain algorithm \cite{BGLL08} from the NetworkKit tool suite \cite{SM16}, and assign $v$ to each of these clusters, which leads to a huge number of overlapping clusters. We therefore delete clusters that are contained in others. Two overlapping clusters are merged\footnote{For the code see \url{https://github.com/ariez-xyz/parallel-merging}} if the second smallest eigenvalue of the normalized Laplacian of the merged cluster is larger than those of the two input clusters by some predefined margin. The clusters to be tested for merging are picked at random, though some heuristics are used to avoid expensive eigenvalue calculations for most of the candidates. Obviously, the resulting clustering ends in a local optimum w.r.t.\ the second smallest eigenvalues of the final clusters. The merging procedure stops when the time needed to find suitable candidates is estimated to be very high. 

\begin{figure}
\centering
\includegraphics[width=0.78\linewidth]{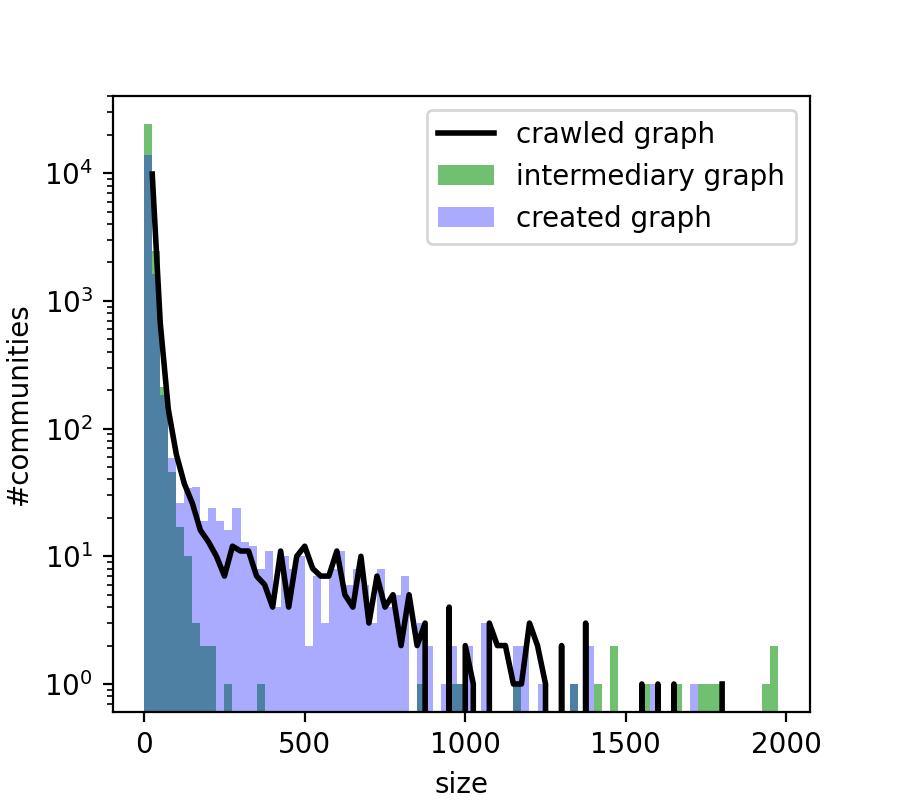}
\caption{Community size distribution in $G_1$}
\label{g1_size_hist}
\end{figure}

In the following, we compare the distribution of the sizes of the overlapping clusters in the crawled graphs, the created graphs, as well as the so-called intermediary graphs, which are constructed as described in Section \ref{METHOD}, but without edge rewiring. We observe that there is no significant difference between the distribution of the cluster sizes in the crawled and the created graphs (see Figures \ref{g1_size_hist} and \ref{g2_size_hist}-\ref{g5_size_hist}; note that for the sake of visualization we use a smoothed distribution curve by averaging each data point with the left and right neighbor for the crawled graphs). This also indicates that the random graphs developed in Section \ref{METHOD} properly model the studied real-world networks. We also compare the distribution of the cluster sizes between the crawled and created graphs on the one side and the intermediary graphs on the other side in Section \ref{sec_intermediary}.

\subsection{Algorithmic Properties}
\label{alg_properties}

One of the major questions in social networks is how certain processes such as information dissemination or epidemics behave in the corresponding graphs. In this section, we compare the behavior of such processes in crawled and created graphs. For information dissemination, we consider the so-called \textit{push-pull} model introduced in \cite{DGHI+87} and popularized later in \cite{KSSV00}. It is known that this model is closely related to certain (structural and algorithmic) graph properties such as the conductance or the cover time of random walks, see e.g., \cite{Gia11}. Note that the computation of these properties is NP-complete in general (cf. \cite{GJ79}). 

The push-pull model is a simple randomized process for information dissemination. The time is divided into rounds; in each round every node selects a neighbor uniformly at random and opens a communication channel to this neighbor. Since the same node can be selected by several neighbors, each node may have several incident communication channels. Then, communication is bidirectional, i.e., if one of the nodes incident to a communication channel possesses some information, the other node will also have this information by the end of the round. At the beginning of the process we place a message on some selected nodes and the question is how many rounds do we need until the message has reached all nodes. Clearly, the number of required rounds is a random variable, with a distribution well concentrated around its mean for many important graph classes \cite{FPRU90}. 

In our experiments, we simulated the push-pull model on the LWCC of the crawled and created graphs. At the beginning, we assigned a message to $2 \log n$ nodes, where $n$ is the number of nodes in the graph, and ran the process until all nodes were informed. Each simulation run has been repeated $100$ times for all graphs except $G_2$ and $G_3$, for which we repeated each simulation $10$ times (as these two graphs are much larger than the others). Although there may be a huge difference between different graph classes w.r.t.\ the completion time of the push-pull model in general \cite{Gia11}, ranging between $\Omega(\log n)$ and $\mathcal{O}(n^2 \log n)$ rounds, we observe that this process behaves similarly on the graphs we consider. 

According to Table \ref{Results_algorithmic}, the number of rounds needed to inform all nodes is lower in the created graphs than in the crawled ones. However, we also observe that the created graphs have a higher density, and they also exhibit a smaller diameter. The crawled graphs also contain a certain number of paths with length $2$ at the periphery\footnote{These paths consist of a node of degree $1$ connected to a node of degree $2$ whose other neighbor is some node of the graph.}, which influence the running time in these graphs, while the number of such paths is much smaller in the created graphs. 

The second process we consider in this section is the so-called \textit{Susceptible-Infected-Recovered} (SIR) model \cite{Het00}. In its simplest variant, we assume that at the beginning a (small) number of nodes is infected. In each round, any infected node spreads the infection to each of its outgoing neighbors with some probability $p$, independently. Then, the nodes that spread the infection in this round become recovered and cannot become infected again. There are several results focusing on the time until the disease dies out as well as on the fraction of nodes becoming infected for different graphs depending on the probability $p$ \cite{Het00,New03}. Since our main focus in this paper is the comparison of the created graphs to the crawled graphs, we do not aim to simulate different variants of the SIR model. We consider the simplest variant instead and compare the influence of the probability $p$ on the number of recovered nodes by the end of the epidemic process. Also in this case, we repeated the simulations on all graphs $100$ times ($10$ times for $G_2$ and $G_3$).

For better comparison, Table \ref{Results_algorithmic} shows the fraction of recovered nodes (instead of the total number of recovered nodes, as the total numbers of nodes differ). We observe that the difference in most of the cases is almost negligible (i.e., less than $0.03$), and the highest difference we observe is less than $0.12$. Overall this simple epidemic process seems to have a similar behavior in the crawled graphs and the corresponding created graphs.

\begin{table}
\small
\caption{Algorithmic properties of $5$ crawled graphs (line 1) and the corresponding created graphs (line 2)}
\centering 
 \begin{tabular}{l c c c c c} 
\toprule
  & $G_1$ & $G_2$ & $G_3$ & $G_4$ & $G_5$\\
  \midrule
  \midrule
  Push-Pull Model \\
  \midrule
  Rounds & 13.92 & 10.64 & 17.24 &  9.79 & 12.40 \\ 
         & 10.78 &  8.95 & 11.22 &  6.30 &  7.62 \\ 
  \midrule
  \midrule
  SIR Model \\
  \midrule
  $p$ = 0.1 & 0.6573 & 0.8812 & 0.8423 & 0.5257 & 0.5317 \\ 
          & 0.6500 & 0.8881 & 0.8098 & 0.6078 & 0.5872 \\
  \midrule
  $p$ = 0.05 & 0.5203 & 0.7994 & 0.7548 & 0.3360 & 0.3755 \\ 
           & 0.5302 & 0.8098 & 0.7248 & 0.3769 & 0.4327 \\
  \midrule
  $p$ = 0.01 & 0.1470 & 0.5015 & 0.4427 & 0.0435 & 0.0370 \\ 
           & 0.1930 & 0.5275 & 0.4560 & 0.0458 & 0.0318 \\
\bottomrule
\end{tabular}
\label{Results_algorithmic}
\end{table}

Motivated by the fact that the behavior of several algorithms is governed by the eigenvalues of the (normalized) Laplacian of the underlying graphs \cite{DFM99,Sau10}, we also analyze the distribution of eigenvalues for the crawled  and created graphs. Similar analysis, using the first $50$ eigenvalues of the corresponding adjacency matrices, has been performed in e.g., \cite{KPPS14}. We observe that there is a difference between the graphs w.r.t.\ the number of eigenvalues close to $1$, while the distribution of the eigenvalues in the ranges $(0,0.95)$ and $(1.05,2)$ are very similar (see Figure \ref{plot_eigenvalues}). The same holds for the other graphs we consider in this paper.

\subsection{Ablation Analysis}
\label{sec_intermediary}

The graph creation method consists of three separate steps, node degree sampling, connecting nodes via edges and rewiring of edges. To demonstrate the importance of correlated degree sampling and edge rewiring, we analyzed the topological and algorithmic features of the created intermediary graphs before applying the rewiring procedure and additionally created a graph with randomly assigned node degrees of the size of $G_1$. 

\subsubsection{Random assignment of node degrees}

The node degrees in our proposed method were sampled using a Copula to match the rank correlations between the degrees in the crawled graphs. To show the importance of joint degree sampling, we created a graph with the same number of nodes as $G_1$, sampled the degrees from the fitted $\chi^2$-distributions, and randomly assigned them to the nodes, similar to the FRD null model in \cite{Durak2013}. As expected, the rank correlations between the degrees in the created graph are around zero, $\rho_1=0.018$, $\rho_2=0.001$, $\rho_3=0.002$. The largest connected components are bigger (LSCC: $8,944$; LWCC: $10,669$), as less nodes have a sampled total degree of zero.

\subsubsection{Results without edge rewiring}

\begin{table}
\small
\caption{Average CC of $5$ created graphs before (*) and after (**) applying the rewiring procedure}
\centering
 \begin{tabular}{l c c c c c} 
\toprule
  & $G_1$ & $G_2$ & $G_3$ & $G_4$ & $G_5$\\ 
 \midrule
 Nodes & 11,015 & 21,291 & 50,133 & 459 & 3,580 \\ 
 
 Average CC* & 0.050 & 0.058 & 0.030 & 0.163 & 0.094\\
 
 Average CC** & 0.228 & 0.291 & 0.222 & 0.300 & 0.187\\
 \bottomrule
\end{tabular}
\label{Results_5_intermediary}
\end{table}

After connecting nodes via edges, the intermediary graphs have node degrees similar to the node degrees in the crawled graphs. We investigated these graphs to assess the influence of the rewiring procedure. While other topological features considered above remain stable, the average CC in the LWCC is significantly lower in all intermediary graphs.
Table \ref{Results_5_intermediary} and Table \ref{Results_before_rewiring} in Annex \ref{Additional_results} contain the average CC of the graphs before (*) and after (**) rewiring. 
Another evidence for the importance of edge rewiring can be observed w.r.t.\ the distribution of community sizes. While the distributions in the crawled and created graphs are similar, the distribution of the community sizes in the intermediary graphs differs significantly from those in the crawled or created ones, cf.~Figure \ref{g1_size_hist}. In contrast, the algorithmic properties analyzed in Section \ref{alg_properties} are similar in the intermediary and the created graphs.

The rewiring procedure takes up most of the graph creation time. Creating graphs without edge rewiring is therefore much faster, especially for bigger graphs, but the results are notably worse.

\subsection{Runtime Analysis}

The experiments were conducted on 1 core of an Intel(R) Xeon(R) 6248 CPU @ 2.50GHz processor with 256GB RAM, and runtimes are listed in Table \ref{Results_5} and in Table \ref{Results_all} in Annex \ref{Additional_results}. It can be seen that the runtime for the graph creation method differs significantly between the graphs. 
While smaller graphs (e.g., $G_1$) are created within $1$ hour, it takes more than a day to create a graph of the size of $G_2$. We observe that the runtime can also vary significantly between graphs with a similar number of nodes (e.g., $G_2$ and $G_7$). This indicates that the runtime depends heavily on the number of edges $|E|$, and in turn on the degree $\deg(x)$ of each node $x \in V'$. 

Since connecting nodes via edges has a fixed time complexity of $\mathcal{O}\left(|V|^2 \right)$ in our algorithm, we conclude that the edge rewiring procedure (Algorithm \ref{Algorithm_rewiring}) dominates the graph creation time. Recall that we are iterating over every node $x$ in the lower $95$-percentile w.r.t.\ their degrees, and check for each pair of its first degree neighbors if there is an edge between these nodes in the graph. If two selected first degree neighbors are connected, they are not further considered in the rewiring process. Otherwise, we look at the nodes which are connected by (incoming and outgoing) incident edges to either of these neighbors, denoted by $y_1$ and $y_2$. Among the neighbors of these two nodes, we do not take into account the nodes with a smaller ID than $x$ as well as those which are connected to both, $y_1$ and $y_2$. From the rest, we select a constant number ($10$ in our implementation) of pairs uniformly at random, and check whether the rewiring procedure described in Algorithm \ref{Algorithm_rewiring} can be applied. 

In order to achieve a space complexity of $\mathcal{O} \left(|V|+|E| \right)$ and runtime $\mathcal{O}(\sum_{x \in V'} \deg(x)^2 \cdot \max(\deg(V')) \cdot \log \max(\deg(V')) )$, where $\max(\deg(V'))$ denotes the maximum degree in $V'$, we need to extend our data structure. We assume that the graph is given as an adjacency list \cite{CLRS01}. Clearly, the list of each node only contains the nodes connected through an outgoing edge in some arbitrary order. In a pre-processing phase, we examine each list entry in the adjacency list exactly once, and construct an additional adjacency list such that the list of a node contains the nodes connected by an incoming edge. Furthermore, for a node $u$ in the outgoing adjacency list of a node $v$ we set a pointer to the node $v$ in the incoming adjacency list of $u$. Before we start the rewiring procedure, we sort each adjacency list according to the IDs of the nodes in this list.
Furthermore, we duplicate each (sorted) adjacency list, and update the duplicates in the rewiring process as follows. Whenever for a chosen node $x$ we finish the rewiring process w.r.t.\ two neighbors $y_1$ and $y_2$ of $x$, then $x$ is removed from the duplicated adjacency lists of $y_1$ and $y_2$. Remember that in the rewiring process $x$ is not considered in any further step whenever $y_1$ or $y_2$ are selected as a first degree neighbor of some node (different from $x$).

Apart from duplicating the adjacency lists, in a step when two nodes $y_1$ and $y_2$ are chosen as first degree neighbors of a node $x$ (see line \ref{Choose} in Algorithm \ref{Algorithm_rewiring}), we first construct two sorted (dynamic) arrays \cite{CLRS01} containing the neighbors of $y_1$ and $y_2$, respectively. Assume now without loss of generality that $\deg(y_1) \leq \deg(y_2)$. We crawl the array of $y_1$ and examine for each neighbor in this array whether it is also connected to $y_2$ (as the adjacency lists of $y_1$ and $y_2$ are sorted, this can be performed for each neighbor of $y_1$ in time $\mathcal{O}\left(\log \deg(y_2)\right)$ using binary search). If this is the case, then the corresponding neighbor of $y_1$ and $y_2$ is marked in the two arrays. After all neighbors of $y_1$ have been examined, the marked nodes are removed from the arrays of $y_1$ and $y_2$. Then, the selection of $10$ random pairs of neighbors of $y_1$ and $y_2$ can be performed in constant time (see line \ref{10-times} of Algorithm \ref{Algorithm_rewiring}). 

Clearly, each node $y_1$ is selected as a first degree neighbor of some other node $\sum_{x \in ne(y_1)} \deg(x) \leq \deg(y_1) \cdot \max(\deg(V'))$ times. Removing the common neighbors of two selected first degree neighbors takes time $\mathcal{O}\left( \deg(y_1) \cdot \log \max(\deg(V')) \right)$. Selecting the second degree neighbors for possible rewiring takes constant time. Sorting the adjacency lists, duplicating them, and creating the arrays takes time $\mathcal{O}\left(\deg(y_1) \cdot \log \max(\deg(V'))\right)$ each time a node $y_1$ is selected as first degree neighbor. Thus, the total time is $\mathcal{O}\left( \sum_{x \in V'} \deg(x)^2 \cdot \max(\deg(V')) \cdot \log \max(\deg(V')) \right)$. Assuming that the average degree in the graph is constant (as for the $\chi^2$ degree distributions we observed in the graphs considered in this paper), we obtain
$\mathcal{O}\left(n \cdot \max(\deg(V'))^2 \cdot \log \max(\deg(V')) \right)$ as runtime.

\section{Discussion and Conclusion}

\label{DISCUSSION}

In this paper we presented a new method to synthetically create simple directed social network graphs with given rank correlations between the reciprocal, in- and out-degrees that can be used as surrogates in social network analysis. We fit $\chi^2$-distributions to the respective degrees of crawled Twitter subgraphs and generate correlated degrees for each node which then connect to other nodes proportional to their degrees. To increase the clustering coefficient of the graphs to a real-world level, we apply an edge rewiring procedure that preserves the node degree sequences.

We compared the created graphs to the crawled ones by analyzing several topological and algorithmic properties. The results show that the graphs are similar regarding the number of edges, the density and the largest connected components. Various topological graph features in the LWCC, e.g., the clustering coefficient, are in the same range. The diameter and the rank correlation between the degrees in the created graphs are slightly lower in all cases. The graphs also show a similar behavior w.r.t.\ the application of information dissemination and epidemics spreading algorithms.

We saw that after the generation of node degrees and the sampling of edges, some nodes have a degree of zero. When applying spreading algorithms, e.g., push-pull or SIR model, these isolated nodes will not be affected if information is only disseminated between neighbors. Since spreading algorithms are usually applied on the largest connected component and not every node is part of it, this circumstance does not seem to be a major concern. Even in the crawled graphs not all nodes are part of the largest connected component, implying that these nodes are not affected by spreading algorithms. Applying a restricted sampling method, where each node ends up with a node degree of at least one, would result in less nodes having a degree of zero, but due to the probabilistic edge sampling, there will still be isolated nodes.

From a theoretical perspective, a limitation of our algorithm is that it does not guarantee unbiased graph generation. That is, there is no guarantee that all random graphs satisfying the considered constraints (degree sequences and their correlation, target range for average CC, etc.) are sampled with the same probability. 
Such a guarantee usually relies on a Markov chain with a high guaranteed mixing time, cf.~\cite{Greenhill15,Fosdick18}, and is therefore not feasible when generating graphs with more than $2 \cdot 10^4$ nodes. Note that several other well-known graph models for real-world networks (e.g., PA or the directed model by~\cite{Bollobas2003}) also induce a certain bias. 

Regarding the efficiency of our graph generation algorithm, the rewiring procedure is the main bottleneck since it takes up most of the runtime, especially for large graphs. We therefore implemented several modifications of the method of \cite{Bansal2009}. The hyperparameters (95-percentile node degree and $60\%$ of first degree neighbor pairs) were determined based on experiments, trading between runtime and outcome. 
Instead of limiting the rewiring procedure to the nodes that have a degree in the lower 95-percentile of neighbors, it could be applied to all nodes (runtime would increase), or fewer nodes, e.g., 90-percentile (runtime would decrease, but the CC would likely be smaller). We also only looked at $60\%$ of first degree neighbor pairs for each node that has a degree above the median degree. Considering more first degree neighbor pairs would lead to a higher CC for these nodes, but increase the runtime.
Parallelizing the rewiring procedure to decrease the runtime does not seem feasible. Each iteration builds on the previous one and the graph constantly evolves with edges being deleted and new edges being created. By looking at multiple nodes at the same time, situations might occur where the same edge is used twice for the rewiring step, leading to errors and altered node degrees.

The results show that the presented approach can be used to create realistic social network graphs that have the same number of nodes and a similar density as crawled subgraphs from the Twitter graph. Our method is highly scalable, i.e., it can generate graphs of arbitrary size with similar properties as real-world networks. We can choose arbitrary rank correlations for the node degrees or select different parameters for the $\chi^2$-distributions. We can also sample the node degrees from other distributions, e.g., a power-law distribution, which has been used to describe the degree distributions in other social network graphs. We are therefore convinced that the proposed approach to create directed social network graphs is a useful alternative to existing network generation methods. 

\begin{acks}
This publication is part of the project "HPC and Big Data Technologies for Global Systems" (HiDALGO), which has received funding from the European Union's Horizon 2020 research and innovation programme under grant agreement No.\ 824115.
The Know-Center is funded within the Austrian COMET Program - Competence Centers for Excellent Technologies - under the auspices of the Austrian Federal Ministry for Climate Action, Environment, Energy, Mobility, Innovation and Technology, the Austrian Federal Ministry for Digital and Economic Affairs and by the State of Styria. COMET is managed by the Austrian Research Promotion Agency FFG.
\end{acks}


\bibliographystyle{ACM-Reference-Format}
\balance
\bibliography{main}

\newpage
\appendix

\section{Additional Results}

\begin{figure}
\centering
\includegraphics[width=0.90\linewidth]{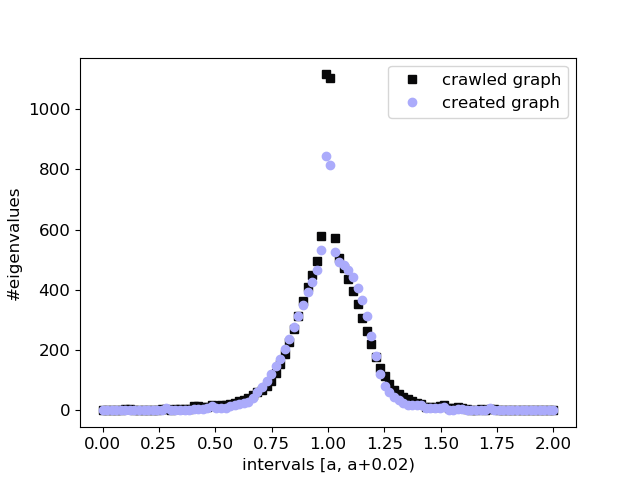}
\caption{Eigenvalue histogram of the normalized Laplacian of the LWCC for $G_1$}
\label{plot_eigenvalues}
\end{figure}

\label{Additional_results}
Table \ref{Results_all} contains the topological features of the $9$ additional graphs we consider, and Table \ref{Results_before_rewiring} contains the average CC before (*) and after (**) applying of the rewiring procedure.
Table \ref{Results_average} contains the averages of the topological features for $4$ created graphs over $5$ runs, which serves as an indicator that our method consistently creates graphs of desired size and properties. The number of edges in the created graphs naturally diverges as we are sampling the reciprocal, the in- and the out-degree from distributions which in turn influences most of the other topological features, but the differences are negligible. 

\begin{figure}[!b]
\centering
\includegraphics[width=0.55\linewidth]{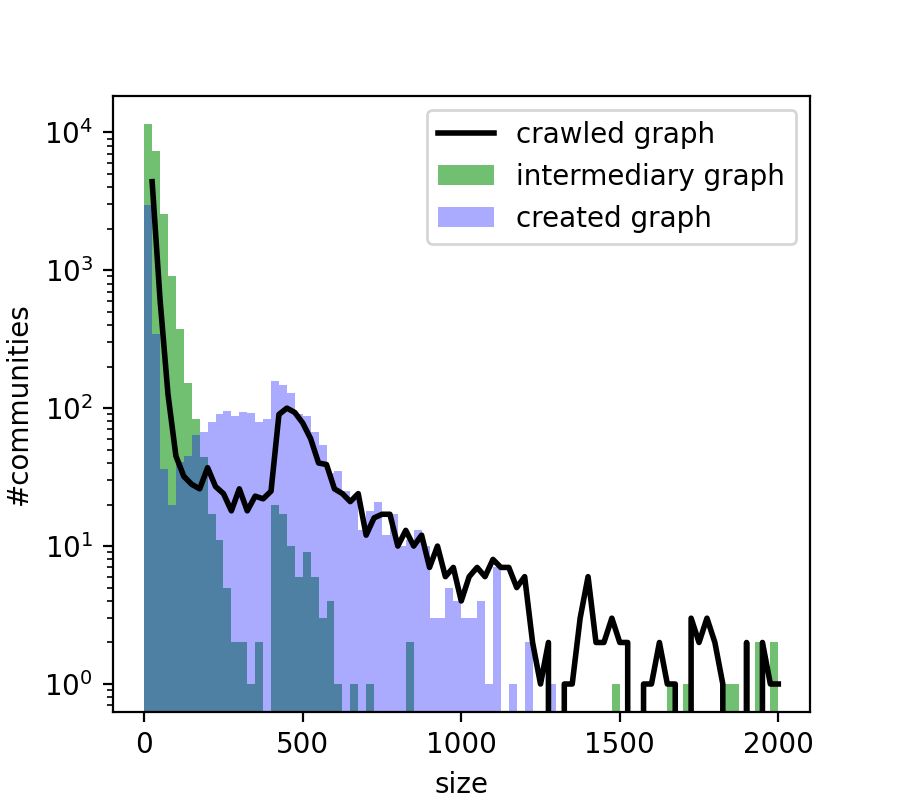}
\caption{Community size distribution in $G_2$}
\label{g2_size_hist}
\end{figure}

\begin{figure}[!b]
\centering
\includegraphics[width=0.55\linewidth]{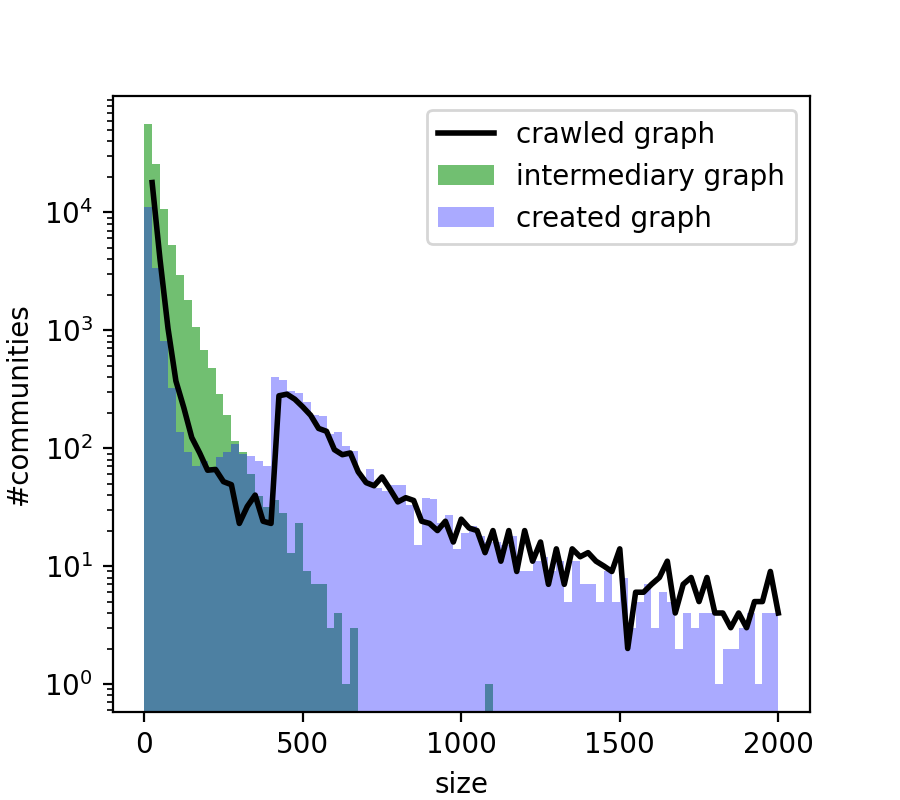}
\caption{Community size distribution in $G_3$}
\label{g3_size_hist}
\end{figure}

\begin{figure}[!b]
\centering
\includegraphics[width=0.55\linewidth]{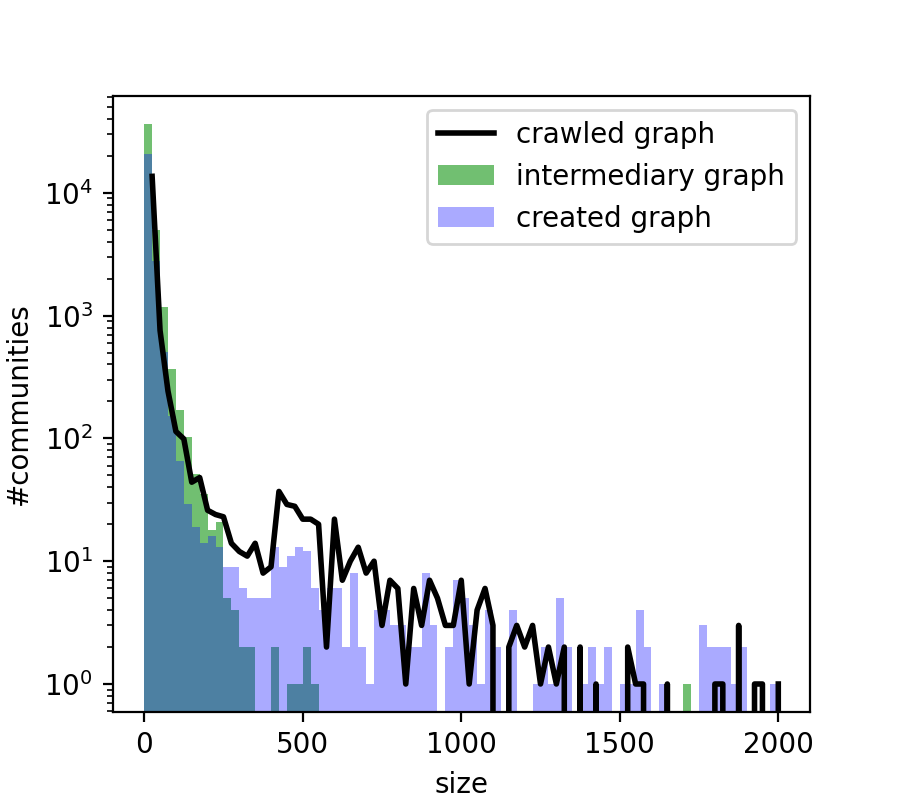}
\caption{Community size distribution in $G_4$}
\label{g4_size_hist}
\end{figure}

\begin{figure}[!b]
\centering
\includegraphics[width=0.55\linewidth]{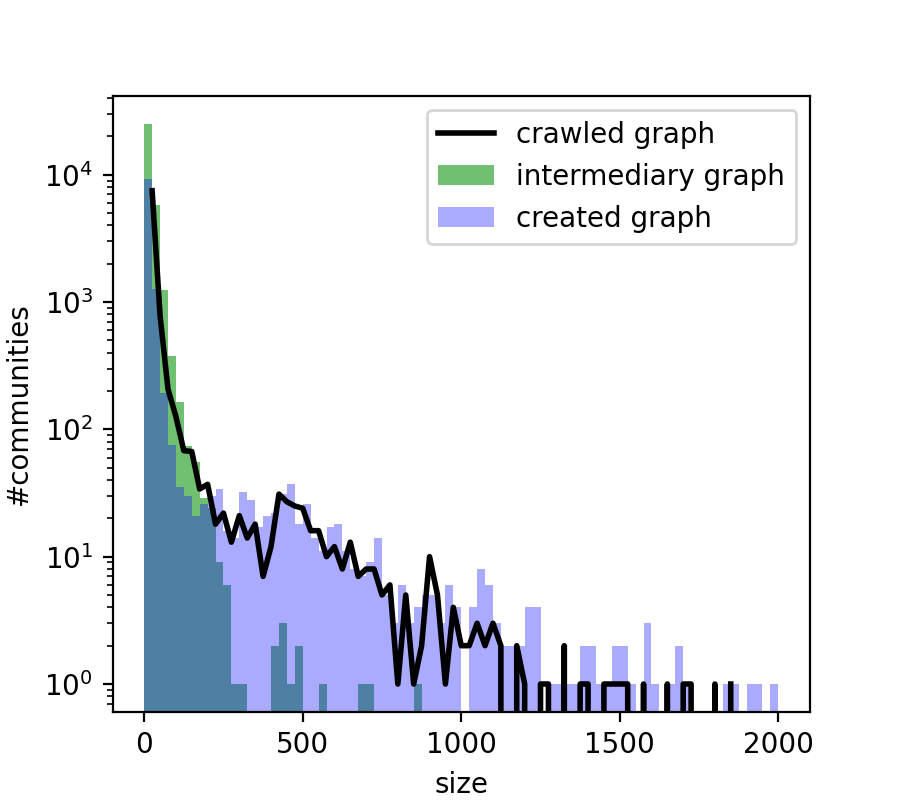}
\caption{Community size distribution in $G_5$}
\label{g5_size_hist}
\end{figure}

Figure \ref{plot_eigenvalues} gives the spectrum of eigenvalues on the LWCC for $G_1$. Similarly to Figure \ref{g1_size_hist}, Figures \ref{g2_size_hist} through \ref{g5_size_hist} reproduce the community size distribution of $G_2$ through $G_5$.

\begin{table*}
\small
\caption{Topological features of the additional crawled (line 1) and the corresponding created graphs (line 2)}
\centering
 \begin{tabular}{l c c c c c c c c c c c c c} 
\toprule
 & $G_6$ & $G_7$ & $G_8$ & $G_9$ & $G_{10}$ & $G_{11}$ & $G_{12}$ & $G_{13}$ & $G_{14}$\\ 
 \midrule
Nodes & 8,277 & 21,464 & 13,646 & 2,013 & 15,299 & 6,003 & 2,464 & 1,239 & 2,932\\
\midrule
Edges & 791,905 & 530,302 & 517,916 & 18,781 & 692,534 & 223,175 & 43,572 & 30,285 & 44,261\\
 & 799,792 & 541,986 & 529,675 & 17,042 & 692,364 & 218,306 & 41,980 & 28,819 & 44,627\\
\midrule
Density & 0.0116 & 0.0012 & 0.0028 & 0.0046 & 0.0030 & 0.0062 & 0.0072 & 0.0197 & 0.0052\\
 & 0.0117 & 0.0012 & 0.0028 & 0.0042 & 0.0030 & 0.0061 & 0.0069 & 0.0188 & 0.0052\\
 \midrule
LSCC & 7,352 & 12,793 & 9,977 & 1,154 & 13,001 & 5,286 & 2,100 & 1,094 & 1,917\\
 & 7,105 & 11,411 & 9,639 & 990 & 12,173 & 5,009 & 1,882 & 1,141 & 1,686\\
 \midrule
LWCC & 8,272 & 21,418 & 13,631 & 1,912 & 15,288 & 5,964 & 2,443 & 1,224 & 2,797\\
 & 8,203 & 19,946 & 13,022 & 1,609 & 14,574 & 5,805 & 2,290 & 1,222 & 2,493\\
\midrule
Density* & 0.0116 & 0.0011 & 0.0028 & 0.0051 & 0.0030 & 0.0063 & 0.0073 & 0.0202 & 0.0056\\
 & 0.0119 & 0.0014 & 0.0031 & 0.0066 & 0.0033 & 0.0065 & 0.0080 & 0.0193 & 0.0072\\
\midrule
ASPL* & 2.52 & 2.49 & 2.69 & 2.65 & 2.65 & 2.70 & 3.04 & 2.68 & 2.41\\
 & 2.33 & 2.00 & 2.38 & 2.11 & 2.64 & 2.68 & 2.55 & 2.59 & 2.18\\
\midrule
Diameter* & 7 & 10 & 10 & 11 & 9 & 9 & 10 & 7 & 11\\
 & 5 & 7 & 7 & 7 & 7 & 7 & 7 & 5 & 7\\
\midrule
Average CC* & 0.339 & 0.266 & 0.244 & 0.260 & 0.319 & 0.234 & 0.321 & 0.303 & 0.310\\
 & 0.380 & 0.150 & 0.215 & 0.176 & 0.240 & 0.240 & 0.251 & 0.312 & 0.210\\
\midrule
Runtime & 9h & 2h & 2h & 1min & 2.5h & 1h & 5min & 2min & 4min\\
\midrule
$\rho_1$ & 0.619 & 0.420 & 0.489 & 0.262 & 0.586 & 0.658 & 0.592 & 0.656 & 0.419\\
 & 0.578 & 0.345 & 0.425 & 0.167 & 0.523 & 0.578 & 0.457 & 0.544 & 0.333\\
\midrule
$\rho_2$ & 0.707 & 0.605 & 0.664 & 0.253 & 0.584 & 0.564 & 0.577 & 0.611 & 0.315\\
 & 0.676 & 0.474 & 0.564 & 0.192 & 0.503 & 0.476 & 0.486 & 0.496 & 0.245\\
\midrule
$\rho_3$ & 0.424 & 0.349 & 0.297 & -0.001 & 0.400 & 0.242 & 0.258 & 0.267 & 0.148\\
 & 0.370 & 0.266 & 0.238 & -0.035 & 0.337 & 0.182 & 0.192 & 0.170 & 0.135\\
\bottomrule
\label{Results_all}
\end{tabular}
\end{table*}

\begin{table*}
\small
\caption{Average CC of the additional created graphs before (*) and after (**) applying the rewiring procedure}
\centering
 \begin{tabular}{l c c c c c c c c c c c c c c c} 
\toprule
  & $G_6$ & $G_7$ & $G_8$ & $G_9$ & $G_{10}$ & $G_{11}$ & $G_{12}$ & $G_{13}$ & $G_{14}$\\
  \midrule
  Nodes & 8,277 & 21,464 & 13,646 & 2,013 & 15,299 & 6,003 & 2,464 & 1,239 & 2,932\\

  Average CC* & 0.122 & 0.040 & 0.053 & 0.073 & 0.049 & 0.060 & 0.083 & 0.113 & 0.093\\

 Average CC**  & 0.380 & 0.150 & 0.215 & 0.176 & 0.240 & 0.240 & 0.251 & 0.312 & 0.210\\
 \bottomrule
\label{Results_before_rewiring}
\end{tabular}
\end{table*}

\begin{table*}
\small
\caption{Average topological features and rank correlations of 4 created graphs over $5$ runs}
\centering
 \begin{tabular}{l c c c c c c c c c c c c c} 
\toprule
  & Nodes & Edges & Density & LSCC & LWCC & Density* & ASPL* & Diameter* & Average CC*  & $\rho_1$ & $\rho_2$ & $\rho_3$\\ 
  \midrule
  $G_1$ & 11,015 & 390,377 & 0.00322 & 8,315.4 & 10,184.2 & 0.00374 & 2.604 & 7 & 0.2306 & 0.4708 & 0.5284 & 0.2354\\
  $G_6$ & 8,277 & 795,898.4 & 0.0116 & 7,118 & 8,189.8 & 0.01188 & 2.334 & 5.2 & 0.376 & 0.5734 & 0.6694 & 0.374\\
   \midrule
  $G_7$ & 21,464 & 522,546.4 & 0.0012 & 11,480.8 & 19,959.6 & 0.0014 & 2.01 & 7 & 0.154 & 0.3486 & 0.4712 & 0.2732\\
  $G_8$ & 13,646 & 526,870 & 0.00282 & 9,670.4 & 13,011.8 & 0.0031 & 2.406 & 7 & 0.2128 & 0.4314 & 0.5698 & 0.2526\\
 \bottomrule
\label{Results_average}
\end{tabular}
\end{table*}

\section{Rewiring Step}

\label{Additional_algorithm}

Algorithm \ref{Algorithm_rewiring_step} depicts the extension of the actual rewiring step from undirected graphs as presented in \cite{Bansal2009} to directed graphs and applied in the rewiring procedure in Algorithm \ref{Algorithm_rewiring}.

\begin{algorithm}[t]
\small
\caption{Rewiring Step}
\label{Algorithm_rewiring_step}
\begin{algorithmic}[1]

\REQUIRE Directed graph $G = (V,E)$

\STATE $y_1, y_2, z_1, z_2 \in V$
\STATE $z_1$ connected to $y_1$, $z_2$ connected to $y_2$

\STATE \textbf{if} $(y_1 \rightarrow z_1) \in E$, $(z_1 \rightarrow y_1) \notin E$, $(y_2 \rightarrow z_2) \notin E$, $(z_2 \rightarrow y_2) \in E$
\STATE \hspace*{3mm} $E = E \setminus \{ (y_1 \rightarrow z_1) \cup (z_2 \rightarrow y_2) \}$ 
\STATE \hspace*{3mm} $E = E \cup (y_1 \rightarrow y_2) \cup (z_2 \rightarrow z_1)$

\STATE \textbf{if} $(y_1 \rightarrow z_1) \notin E$, $(z_1 \rightarrow y_1) \in E$, $(y_2 \rightarrow z_2) \in E$, $(z_2 \rightarrow y_2) \notin E$
\STATE \hspace*{3mm} $E = E \setminus \{ (z_1 \rightarrow y_1) \cup (y_2 \rightarrow z_2) \}$ 
\STATE \hspace*{3mm} $E = E \cup (y_2 \rightarrow y_1) \cup (z_1 \rightarrow z_2)$

\STATE \textbf{if} $(y_1 \rightarrow z_1) \in E$, $(z_1 \rightarrow y_1) \in E$, $(y_2 \rightarrow z_2) \in E$, $(z_2 \rightarrow y_2) \in E$
\STATE \hspace*{3mm} $E = E \setminus \{ (y_1 \rightarrow z_1) \cup (z_1 \rightarrow y_1) \cup (y_2 \rightarrow z_2) \cup (z_2 \rightarrow y_2) \}$ 
\STATE \hspace*{3mm} $E = E \cup (y_1 \rightarrow y_2) \cup (y_2 \rightarrow y_1) \cup (z_1 \rightarrow z_2) \cup (z_2 \rightarrow z_1)$

\end{algorithmic}
\end{algorithm}

\end{document}